\documentclass{article}

\usepackage{bm,amssymb,amsfonts,amsmath}
\usepackage[dvips]{graphicx}
\usepackage[dvips]{color}

\begin{document}

%%%%%%%%%%%%%%%%%%%%%%%%%%%%%%

% \conflictofinterest{No conflict of interests}
% \track{This paper was submitted directly to the PNAS office}

\title{Non-linear dynamics of absorption and photo-bleaching of dyes.}

\author{Francesca Serra and Eugene M. Terentjev \\
{\small Cavendish Laboratory, University of Cambridge, Cambridge,
CB3 0HE, UK}}

%% The \maketitle command is necessary to build the title page.
\maketitle

%%%%%%%%%%%%%%%%%%%%%%%%%%%%%%%%%%%%%%%%%%%%%%%%%%%%%%%%%%%%%%%%
%\begin{article}

\begin{abstract} The celebrated Lambert-Beer law of light
absorption in photochromic media is only valid at low intensities
of incident light and low concentration of chromophore. Here we
address the generic problem of photo-absorption dynamics,
experimentally studying the case of azobenzene isomerization. We
show that the non-linear regime is very common and easy to achieve
in many practical situations, especially in thick samples where
the light depletes the chromophore in the first layers and can
propagate through the medium with a sub-exponential law.
Importantly, the crossover into the non-linear absorption regime
only weakly depends on the dye concentration and solution
viscosity. We experimentally quantify the characteristics of this
peculiar optical response and determine the key transition rate
constants. \end{abstract}

%% When adding keywords, separate each term with a straight line: |
\noindent {\sf Keywords: Lambert-Beer law | photobleaching |
azobenzene | isomerization kinetics | absorption spectroscopy | }

%% The first letter of the article should be drop cap: \dropcap{}
\subsection*{Introduction}
Lambert-Beer law is well known and widely used in spectroscopy: it
states that the light propagating in a thick absorbing sample is
attenuated at a constant rate, that is, every layer absorbs the
same proportion of light \cite{jaffe}. This can be expressed in a
simple form as the remaining light intensity at a depth $x$ into
the sample: $I(x) = I_0 \exp(-x/D)$ where $I_0$ is the incident
intensity and $D$ is a characteristic length which one calls the
``penetration depth'' of a given material. If an absorbing dye is
dispersed in a solution (or in an isotropic solid matrix) this
penetration depth is inversely proportional to the dye
concentration. In this way it is possible to determine a dye
concentration $c$ by experimentally measuring the absorbance,
defined as the logarithm of intensity ratio $A = \ln
\big(I_0/I\big)= x(c/\delta)$, where $x$ is the thickness of the
sample (the light path length), $c$ the concentration of the
chromophore, and $\delta$ the universal length scale
characteristic of a specific molecule/solvent. Thanks to
Lambert-Beer law, UV-visible absorption spectroscopy is a useful
and practical tool in many areas of science, for instance, in
biology to determine the degree of purity of a protein or the
concentration of DNA \cite{Zaccai}. One should note that in
chemistry and biology one often uses base-10 logarithm in defining
the Absorbance, $A_{10}= \log I_0/I = (\varepsilon \, c)x$. If $c$
is in molar units, the constant of proportionality $\varepsilon$
is called the ``molar absorption coefficient'' and it is inversely
proportional to the characteristic length $\delta$ defined above.

However, this empirical law has limitations, and deviations are
observed due to aggregation phenomena or electrostatic
interactions between particles. More importantly, in
photosensitive media much stronger deviations due to self-induced
transparency, or photobleaching \cite{hahn,armstrong} can occur.
This effect has been reported in many biological systems like
rhodopsin \cite{yoshizawa, merbs}, green fluorescent protein
\cite{henderson} or light harvesting complexes \cite{bopp}.

 In most photosensitive molecules irradiation with light
at a certain wavelength induces a conformational change
(isomerization) from an equilibrium  \textit{trans} state where the
benzene rings are far apart to a bent \textit{cis} state where they
are closer. The particular characteristic that makes azobenzene and
its derivatives interesting is that the two isomers have different
absorption spectra, primary absorption peaks in two different
regions: the \textit{trans} isomer absorbs around 320 nm, while the
\textit{cis} isomer at around 440 nm \cite{Rau}. The position of the
two peaks is sensitively changed by a variety of chemical groups
attached to the basic azobenzene molecules, making azobenzene
derivatives family a model material for study of many photo-chromic
phenomena. Irradiation with light at the wavelength of the
\textit{trans} peak progressively depletes the molecules in this
conformation. Effective \textit{trans-cis} photo-bleaching of the
first layers allows a further propagation of light into the sample
and this leads to nonlinear phenomena which are interesting both
from the theoretical \cite{andorn,berglund,janossi,corbett} and from
the experimental point of view
\cite{meitzner,barrett,broer,vanoosten}.

%\section{Theoretical background}

The non-Lambertian propagation of light through a medium has
important consequences for the analysis of photo-isomerization
kinetics: when the photo-bleaching becomes important, the measured
absorbance does not follow a simple (traditionally used)
exponential law anymore. The isomerization process follows the
first-order kinetics:
\begin{equation}
\frac{dn_t}{dt}=-I_{TC}k_{TC}n_t+I_{CT}k_{CT}(1-n_t)+\gamma
(1-n_t) \label{kin1}
 \end{equation}
 where
$n_t$ is the fraction of isomers in the \textit{trans} state,
 $k_{CT}$ and $k_{TC}$ are the
\textit{cis-trans} and \textit{trans-cis} constants of
photoisomerization, respectively, $I_{TC}$ and $I_{CT}$ are the
intensities of light at the wavelengths which excite the two
transitions, and $\gamma$ is the rate of spontaneous thermal
\textit{cis-trans} isomerization. In the experiments described
below we use an azobenzene derivative in which the
\textit{trans-cis} and \textit{cis-trans} absorption peaks are
widely separated and illuminating light monochromated at the
\textit{trans-cis} transition wavelength. In this case the
stimulated \textit{cis-trans} isomerization is negligible (that
is, $I_{CT} \rightarrow 0$) and the kinetic equation reduces to
\begin{equation} \frac{dn_t}{dt}=-\gamma\Big( \left[1+Ik_{TC}/\gamma
 \right] n_t+1\Big) . \label{kin2}
 \end{equation}
In this equation the intensity $I=I(x)$, depending on the depth
into the sample. It is convenient to define a non-dimensional
parameter $\alpha= I_0 k_{TC}/\gamma$,  which represent the
balance of photo- and thermal isomerization at a given incident
intensity $I_0$. In this notation, the amount of molecules in the
\textit{trans} conformation in the photostationary state is simply
\begin{equation}
n_t=(1+\alpha I/I_0)^{-1} . \label{ent}
 \end{equation}

\begin{figure} %[!h!]
\centering
\includegraphics[width=0.96\textwidth]{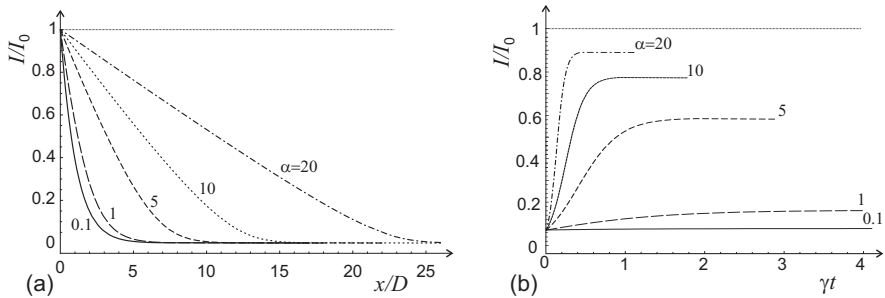}
\caption{(a) Intensity $I/I_0$ in the photostationary state as a
function of the sample thickness $x/D$ for several values of
$\alpha$. At small $\alpha$ the decay is exponential; the light
penetrates deep into the sample as $\alpha$ increases. \ (b)
Intensity $I/I_0$ as function of time through a fixed $x/D=2.7$
and different incident intensity; as $\alpha$ grows, the dye
reaches the stationary state increasingly fast with a behavior
which differs from a simple exponential.
  \label{stationary} }
\end{figure}

The change in intensity between subsequent layers of the absorbing
material is proportional to the conversion of \textit{trans}
molecules in the layer. This theoretical problem has been studied
by Statman and Janossy \cite{janossi}, and Corbett and Warner
\cite{corbett}. Neglecting the stimulated \textit{cis-trans}
isomerization (which is appropriate in our study, the model can be
much simplified to give:
\begin{equation}
\frac{dI}{dx}=-I \frac{n_t}{D} \label{intvar}
 \end{equation}
where \textit{D} is the penetration depth, inversely proportional
to concentration. Combining (\ref{intvar}) and (\ref{ent}) the
stationary-state light intensity at a depth $x$ is given by the
relationship  \cite{corbett}
\begin{equation}
\ln\big[I(x)/I_0\big]+\alpha \big[ I(x)/I_0-1\big]=-x/D .
\label{photostat}
 \end{equation}
Figure \ref{stationary}(a) shows the intensity variation $I(x)$
for various values of the parameter $\alpha$: if the incident
intensity is low enough ($I_0 k_{TC}/\gamma \rightarrow 0$), the
Lambert-Beer law is valid and the decrease is exponential, but if
the incident intensity is high the bleaching of the first layers
becomes progressively relevant so that they become partially
transparent to the radiation. The decay thus tends to become
linear in the large part of the sample, $I(x) \approx I_0
(1-x/\alpha D)$.

In order to model the dynamics of photoisomerization, which is
evidently inhomogeneous across the sample, equations (\ref{kin2})
and (\ref{intvar}) must be coupled to give
\begin{equation}
\frac{d}{dt}\left(\frac{1}{I}\frac{dI}{dx}\right)=
-\frac{1}{D}\frac{dn_t}{dt}=-\frac{\gamma}{D}\left(1-\left[\alpha
I/I_0 +1\right] n_t\right) \label{intkin}
 \end{equation}
Solving this equation (see \cite{corbett2} for detail) leads to
the integral expression for the intensity $I(x,t)$ that we would
detect from the sample of thickness $x$ at time $t$:
\begin{equation}
 \gamma t= - \int_{x/D}^{A}\frac{dy}{\big(x/D-y -\alpha +\alpha
e^{-y} \big)} . \label{kinfin}
 \end{equation}
The upper limit of this integral is the measurable absorbance
$A=\ln [I_0/I(x,t)]$ from a sample of thickness $x$. Figure
\ref{stationary}(b) shows the resulting prediction for the
time-evolution of intensity transmitted along the path $x/D$. Note
that at $t=0$ all curves converge to the Lambert-Beer $I/I_0 =
\exp [-x/D]$, while at long times a significant portion of
chromophore is bleached and transmitted intensity increases.

We should note that, however important the practical case of
molecular isomerization might be, the problem of non-linear
photo-absorption dynamics is much more generic. Even ordinary dye
molecules that do not undergo conformational changes stimulated by
photon absorption, still follow the same dynamic principles, only
with electronic transitions in place of  \textit{trans-cis}
isomerization. Therefore, the results of this paper should be
looked upon in such a generic sense. In particular, the two key
conclusions: that the crossover intensity into the non-linear
photo-absorption regime is independent on dye concentration and
the rate of the transition is independent on solvent viscosity,
are probably completely general.

\subsection*{Materials and Methods} 
\begin{figure} [!h!]
\centering
  \includegraphics[width=0.45\textwidth]{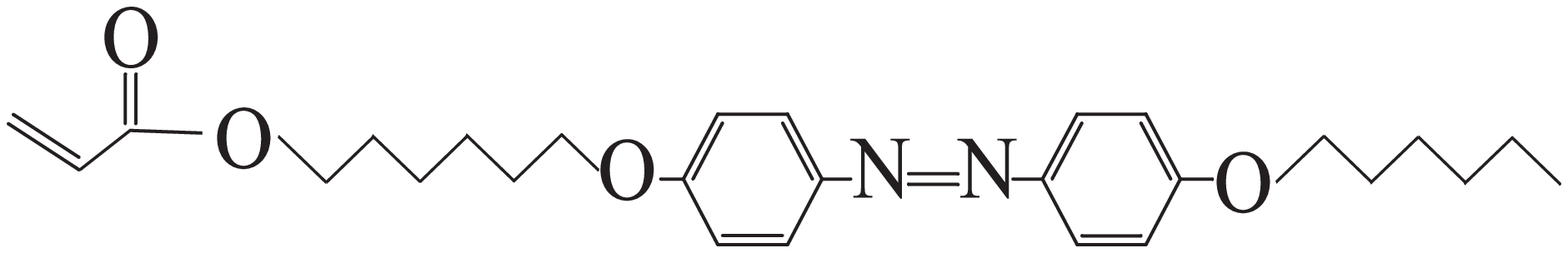} \\
   \caption{The monomer AC$_6$AzoC$_6$ has the azobenzene group
  symmetrically positioned between two similar aliphatic chains,
  one of which is terminated by an acrylate group.
  \label{acry} }
\end{figure}

The photochromic derivative
4'-hexyloxy-4-((acryloyloxy)hexyloxy)-azobenzene (abbreviated as
AC$_6$AzoC$_6$) was synthesized in our lab by Dr. A.R. Tajbakhsh.
The molecular structure is shown in figure \ref{acry} and the
synthesis described in \cite{me}. The absorption spectrum of the
molecule has a peak centered at 365 nm, characteristic of
the\textit{ trans }state, and a peak at 440 nm characteristic of
the \textit{cis} state. We deliberately choose this molecule with
widely separated absorption peaks to be able to unambiguously
monitor the kinetics of \textit{trans-cis} transition.

The illumination was provided by a Nichia chip-type UV-LED, whose
output power was accurately regulated by a power supply. The
spectrum of emission of the light is centered at 365 nm (near the
absorption maximum of our azobenzene derivative) and is about 10
nm wide. The LED light was attenuated by passing through a black
tube of controlled length, placed in front of a quartz cuvette
with 1 cm optical path. We determined the power density of light
incident on the sample by calibration against the LED input
current $J~\hbox{[mA]}$, the illumination cone angle of the LED
(132${}^{\circ}$) and the illuminated area at the fixed distance
to the sample (20~cm), giving the intensity $I_0 = 0.4J~\mu
\hbox{W/cm}^2$. Several values of intensity were used in reported
measurements, ranging between $I_0 = 4$ and $60~\mu
\hbox{W/cm}^2$.

For light absorbtion measurements of a Thermo-Oriel MS260i
spectrometer (focal length 260mm) was used.  A shutter was placed
in front of the lamp so that the sample can be rapidly screened
from the light source. An liquid lightguide conducted the light
from the cuvette to the spectrometer, through a 50 $\mu$m slit and
the Andor linear-array CCD camera connected to a computer. The
camera allowed a rate as fast as 0.021 s per frame. Before every
absorption experiment, a background and a reference spectra were
collected: the background is the spectrum without the
illumination, and the reference is the spectrum collected with the
LED illuminating the cuvette filled with plain solvent without the
chromophore. The absorbance is then calculated from the counts of
the detector as:
 $$A=\ln \Bigg(\frac{\rm Reference-Background}{\rm Signal-Background}\Bigg).$$
All experiments were carried out at room temperature in the dark.

\begin{figure} %[!h]
\centering
 \includegraphics[width=0.96\textwidth]{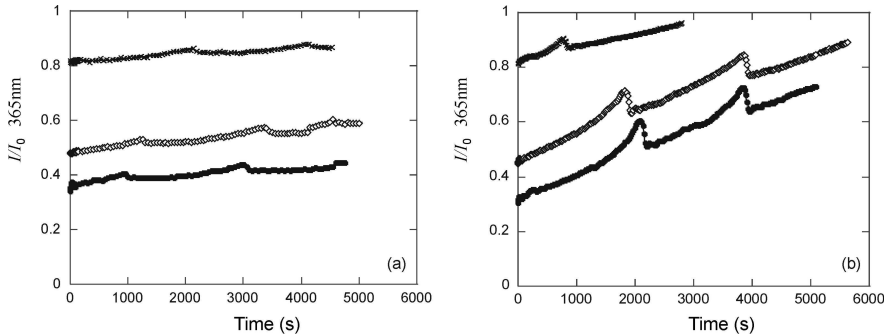}
 \caption{$I/I_0$ for 3 different values of $x/D$ ($\times$ - 0.2,
$\diamondsuit$ - 0.7, $\bullet$ - 1.1) and 2 different values of
$\alpha$, corresponding to: (a) $I_0= 4 \mu \hbox{W cm}^{-2}$, and
(b) $I_0= 20 \mu \hbox{W cm}^{-2}$. The periodic instability was
reproducible in all low-viscosity experiments. \label{conv}}
\end{figure}

For our detailed dynamic experiments, a very important issue was
the viscosity of the solution. In the earlier work \cite{me} we
have shown that solvent viscosity has no significant effect on the
\textit{trans-cis} transition rate. However, at high illumination
intensity we have encountered an unexpected problem. Figure
\ref{conv} shows that the transmission of light through a
low-viscosity dye solution (in pure toluene) displays a
characteristic oscillatory feature. Detailed analysis of this
phenomenon is beyond the scope of this paper. Whether the
oscillations are linked to the local convection due to the heating
of the sample spot \cite{nitzan} or to the diffusion of the less
dense \textit{cis} molecules -- or whether they are intrinsic to
the non-linear photochemical process \cite{borderie} -- is not
clear at this stage and would need further investigation.

In order to avoid this difficulty, the dye solutions were prepared
in a mixture of toluene and polystyrene of high molecular weight
(Mw=280000), both purchased from Sigma-Aldrich and used without
further purification. Adding polystyrene increases the viscosity of
the solution by over 2 orders of magnitude, and in this way prevents
fluid motion in the cuvette on the time scales of our measurements.
Polystyrene-toluene solutions were prepared at a fixed weight ratio
of 37.5wt\%.

In all cases it was important to verify that the dye concentration
remained in a range where, at time $t=0$, the linear
proportionality between absorbance and concentration (Lambert-Beer
law) held. This is important because the concentration of
molecules in the \textit{trans} state at every instant was
determined from the absorbance at 365 nm. Absorbance was measured
at several concentrations. The deviation from linearity started at
$A \approx 3$, which corresponds to the dye weight fraction of
$c=0.03$ (3 wt\%) in our 1 cm cuvette. After this point,
aggregation effects start playing a role and the basic
Lambert-Beer law is no longer valid, undermining the theoretical
relationship given by the equation (\ref{kinfin}). We always kept
the concentrations below this value, so that the linearity at
$t=0$ was maintained, with $A=x/D$ where the penetration depth is
inversely proportional to chromophore concentration, expressed as
weight fraction, $D \approx \delta/c$ with $\delta = 91\, \mu$m.

We prepared three different dye solutions with (non-dimensional)
weight fractions $c=2.5 \cdot 10^{-3}$, $0.01$ and $0.025$,
resulting in the values of penetration depth $D=36$~mm, $9.1$~mm
and $3.6$~mm, respectively (the sample thickness was kept constant
at $x=10$~mm).

Every point of the spectrum was collected as an average of 100
measurements. All isomerization reactions were followed for
several hours, until a photostationary state was reached. The
measurements were repeated at different illumination intensities
(regulated with the power supply) and at different dye
concentrations.

\subsection*{Results and discussion}

A basic characteristic of the photoisomerization problem is the
rate of spontaneous thermal \textit{cis-trans} isomerization
$\gamma$. For a given azobenzene derivative, at fixed (room)
temperature and sufficiently low dye concentrations to avoid
self-interaction, this rate is approximately the same for all our
solutions. We measured this rate after monitoring the relaxation
of the spectrum after the illuminating LED is switched off (see
\cite{me} for detail) and obtained $\gamma \approx 1.25 \cdot
10^{-4} \hbox{s}^{-1}$ (or the corresponding relaxation time of
$\sim 8000$s).

In order to test the predictions of the theory, dynamic absorption
measurements were performed for different dye concentrations and
different light intensities. Taking the equation (\ref{kinfin})
this is equivalent to changing $x/D$ (where $D$ is inversely
proportional to the dye concentration) and $\alpha$, which is
proportional to the incident intensity $I_0$. With this
experimental setup it was possible  to follow all the
isomerization kinetics and thus the time dependence of $I/I_0$.

\begin{figure} %[!h!]
\centering
  \includegraphics[width=0.52\textwidth]{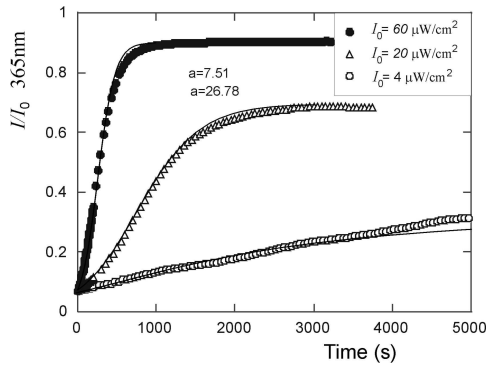}
  \caption{The effect of photo-bleaching for samples with high dye
  concentration ($x/D=2.7$). Three values of irradiation intensity
  are labelled on the plot. Solid lines are fits to the data with
  only one free parameter $\alpha$, giving $\alpha \approx 24.7$
  for the highest intensity, and $\alpha \approx 7.5$ for the
  middle intensity.
  \label{highX} }
\end{figure}

At time $t=0$ the absorption follows the Lambert-Beer law, which
guarantees the proportionality between the absorbance and the dye
concentration.  In figure \ref{highX} the representative
experimental results are shown for the solution with the highest
chromophore concentration ($D=3.6$~mm,leading to $x/D=2.7$) and
three values of incident intensity. One finds that all curves
converge to the same initial value corresponding to the
$I/I_0=\exp [-x/D]$, which for this concentration means quite a
low transmission ($I/I_0 \approx 0.07$). The classical
Lambert-Beer would correspond to this value remaining
time-independent. We fit the data with the theoretical model given
by equation (\ref{kinfin}) where we input the values of $\gamma$
and $x/D$, leaving only $\alpha = I_0 k_{TC}/\gamma$ free. Two
data sets at higher intensity show the transmitted $I(x,t)$ reach
saturated value. In these case the fit is confident because we
have to match both the slope and the amplitude of the curve. We
obtain $\alpha \approx 24.7$ for $I_0 = 60~\mu \hbox{W/cm}^2$ and
$\alpha \approx 7.5$ for $I_0 = 20~\mu \hbox{W/cm}^2$, the ratio
of which is close to expected $3$. For the low-intensity
irradiation, although the deviation from the Lambert-Beer law is
evident, our fitting is not reliable because we did not obtain the
intensity saturation in a reasonable time of experiment.

\begin{figure} %[!h!]
\centering
  \includegraphics[width=0.52\textwidth]{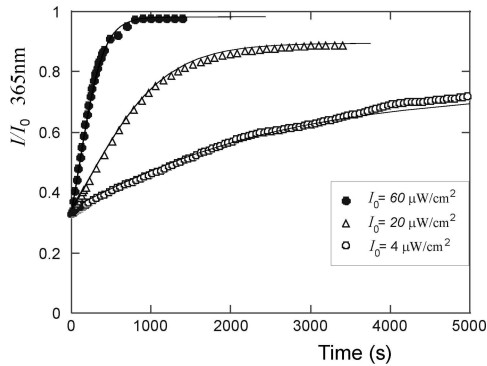}
  \caption{The same experiment as in figure \ref{highX} but with an intermediate
dye concentration ($x/D=1.1$), and the same values of irradiation
intensity. Here the solid lines are not fits, but theoretical
plots of equation (\ref{kinfin}) for $\alpha = 24.7, 7.5$ and
$1.64$ for the decreasing intensity, respectively.
  \label{midX} }
\end{figure}

At lower concentration of chromophore, corresponding to $D\approx
9.2$~mm and $x/D=1.1$, figure \ref{midX} shows the similar
features of the non-linear effect. Only at higher irradiation
intensities we could achieve the saturation and the steady-state
value $I(x)$ corresponding to the solution of equation
(\ref{photostat}). The change of curvature, notable in figures
\ref{stationary})(b) and \ref{highX}, is not so clear here even at
the highest $I_0$. However, in the comparative analysis of data we
now take a different approach. Assuming all the parameters for the
curves are now known ($\gamma$ and $x/D$ from independent
measurements, and $\alpha$ from the fitting in figure
\ref{highX}), we simply plot the theoretical equation
(\ref{kinfin}) on top of the experimental data. For the
low-intensity curve we take $\alpha = 1.64$, which is a 1/5
fraction of the value for the highest $I_0$ assuming the
linearity. It is clear that the theory is in excellent agreement
with the data.

\begin{figure} %[!h!]
\centering
  \includegraphics[width=0.52\textwidth]{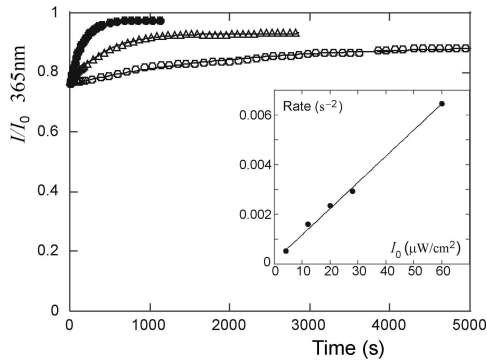}
  \caption{At low dye concentration ($x/D=0.27$) the sample
  is relatively transparent. The data are for the same three
  values of irradiation intensity as in the earlier plots.
The inset shows the plot of exponential relaxation rate
$\tau^{-1}$ against $I_0$, with the linear fit. \label{lowX} }
\end{figure}

Finally, we examine the dynamics of non-linear absorption at low
dye concentration ($D\approx 91$~mm, $x/D=0.27$) in figure
\ref{lowX}.  In this case the complicated integral equation
(\ref{kinfin}) simplifies dramatically \cite{corbett2}, since at
small $x/D \ll 1$ the difference between $A=\ln [I_0/I] $ and
$x/D$ (which is the range of integration in (\ref{kinfin}), is
also small. The integration can then be carried out analytically,
giving
\begin{equation}
\ln \left( \frac{I(x,t)}{I_0} \right) \approx -\frac{x}{D} \left[
1-\frac{\alpha}{1+\alpha} \left( 1- e^{-\gamma(1+\alpha)t} \right)
\right] , \label{approx1}
\end{equation}
which gives in the stationary state the correct solution of
equation (\ref{photostat}) approximated at small $x/D$:
$$\ln \left( \frac{I(x)}{I_0} \right) \approx
-\frac{x/D}{1+\alpha}.
 $$
More importantly, we also see that that the rate of the process
described by the approximation (\ref{approx1}) is given by the
simple exponential, $\tau^{-1} = \gamma(1+\alpha) = \gamma +
k_{TC}I_0$. This is in fact the rate originally seen in the
kinetic equation (\ref{kin2}). Therefore, we now fit the family of
experimental curves in figure \ref{lowX} by the simple exponential
growth and then analyze the relaxation rates obtained by this fit.
The inset in figure \ref{lowX} plots these rates for all the $I_0$
values we have studied. A clear linear relation between the
relaxation rate and $I_0$ allows us to independently determine the
molecular constant: $k_{TC} \approx 10^{-4} \hbox{cm}^2
\hbox{s}^{-1} \mu \hbox{W}^{-1}$.

\subsection*{Conclusions} The main conclusion of this paper is that
one has to be cautious with the classical concept of light
absorption, represented by the Lambert-Beer law. The question is
not about more complicated multi-particle effects at high
concentration of photo-sensitive molecules in the system:
obviously for very high absorbance values one does not expect
linearity. Importantly, even at very low concentrations
(corresponding in our case to the low $x/D$ ratio) the
illumination intensity above a certain crossover level would
always produce a non-linear dynamical effect equivalent to the
effective photo-bleaching, increasing the effective transmittance
of the sample. The crossover is expressed by the non-dimensional
parameter $\alpha = I_0 k_{TC} /\gamma \geq 1$ and is, therefore,
an intrinsic material parameter of every chromophore molecule, but
not dependent on the dye concentration. Note that the thermal
\textit{cis-trans} isomerization rate $\gamma$ is strongly
temperature dependent, influencing the crossover intensity.

The experiments, carried out to quantify this dynamical effect,
have confirmed the model predictions and demonstrate several
linked  features, especially pronounced in the high-concentration
systems. The final saturation steady-state value of transmitted
intensity (or the practically measured absorbance) is linked with
the rate at which the steady state is reached. At high intensity
of illumination one finds a characteristic sigmoidal shape, with
the changing curvature of $I(t)$ variation. Our experiments were
deliberately carried out in a highly viscous solvent to eliminate
the additional complexities caused by the possible local
convection flows of different isomers, which are illustrated in
figure~\ref{conv}. Certainly, a much more in-depth study will be
required to take such effects into account.

It is important to emphasize that these results and conclusions
are very general, not at all confined to the class of azobenzene
or other photo-isomerizing molecules. All dyes are essentially
systems with two long-lived energy levels and thus the non-linear
kinetics of their absorption and saturation will obey the same
rules as discussed in this paper.

%\begin{acknowledgments}
\subsubsection*{Acknowledgments}
We acknowledge the Cambridge Nanoscience Centre for allowing
access to the Oriel fast spectrometer and A.R. Tajbakhsh for
synthesizing the azobenzene derivative used in this study. Useful
discussions with D. Corbett, J. Huppert and M. Warner are
gratefully appreciated. This work has been supported by EPSRC and
Mars U.K.
%\end{acknowledgments}

% Sample bibliography item in PNAS format:
%% \bibitem{Ch} D. Chae (2003) {\it Nonlinearity} {\bf 16}, 479-495.
%% \bibitem{in-text reference} Author Names (year published)
%% {\it Journal Name} {\bf Volume #}, start page-end page

%bibliography new style as suggested by the website

%\end{document}
%%%%%%%%%%%%%%%%%%%%%%%%%%%%%%%%%%%%%%%%%%%%%%%%%%%%%%%%%%%%%%%%
\end{document}